\def\citenum#1{{\def\@cite##1##2{##1}\cite{#1}}}

\def\spose#1{\hbox to 0pt{#1\hss}}
\def\lsim{\mathrel{\spose{\lower 3pt\hbox{$\mathchar"218$}}
     \raise 2.0pt\hbox{$\mathchar"13C$}}}
\def\gsim{\mathrel{\spose{\lower 3pt\hbox{$\mathchar"218$}}
     \raise 2.0pt\hbox{$\mathchar"13E$}}}
\def\simpropto{\mathrel{\spose{\lower 3pt\hbox{$\mathchar"218$}}
     \raise 2.0pt\hbox{$\propto$}}}

\def\TB{T_{\slash \!\!\!\! B}}
\def\rs{\slash \!\!\!\! R}

\def\hc{\rm H.c.}

\def\eg{ {\it e.g.} }

\def\PRL#1#2#3{{\sl Phys. Rev. Lett.} {\bf #1}, #2 (#3)}
\def\PRD#1#2#3{{\sl Phys. Rev.} {\bf D#1}, #2 (#3)}
\def\PLB#1#2#3{{\sl Phys. Lett.} {\bf B#1}, #2 (#3)}
\def\PREP#1#2#3{{\sl Phys. Rep.} {\bf #1}, #2 (#3)}
\def\NPB#1#2#3{{\sl Nucl. Phys.} {\bf B#1}, #2 (#3)}

\def\tev{{\rm TeV }}
\def\gev{{\rm GeV }}
\def\Mev{{\rm MeV }}

\def\msusy{M_{\rm SUSY}}

\def\mz{m_{\rm z}}

\def\half{{1\over 2}}

\def\ifmath#1{\relax\ifmmode #1\else $#1$\fi}

\def\be{\begin{equation}}
\def\ee{\end{equation}}
\def\bea{\begin{eqnarray}}
\def\eea{\end{eqnarray}}
\def\bq{\begin{quote}}
\def\eq{\end{quote}}

\def \lsim{\mathrel{\vcenter
     {\hbox{$<$}\nointerlineskip\hbox{$\sim$}}}}
\def \gsim{\mathrel{\vcenter
     {\hbox{$>$}\nointerlineskip\hbox{$\sim$}}}}

\def\gappeq{\mathrel{\rlap {\raise.5ex\hbox{$>$}}
{\lower.5ex\hbox{$\sim$}}}}

\def\lappeq{\mathrel{\rlap{\raise.5ex\hbox{$<$}}
{\lower.5ex\hbox{$\sim$}}}}

\def\bbz{fa Z \kern-8.9pt Z}

\documentstyle [12pt]{article}
\begin{document}
\thispagestyle{empty}
\begin{flushright}
{UCDPHY-96-27} \\
{MPI-PhT-96-89} \\
{September 1996} \\
\end{flushright}
\vspace{1cm}
\begin{center}
{\large  Protecting the Baryon Asymmetry in Theories
with R-parity Violation} \\
\end{center}
\vspace{1cm}
\begin{center}
{Sacha Davidson$^1$ and Ralf Hempfling$^{1,2}$ }\\
\vspace{.3cm}
{$^1$Max Planck Institut f\"{u}r Physik\\
F\"{o}hringer Ring 6, D-80805, M\"{u}nchen, Germany}\\
\vspace{.3cm}
{$^2$Univ. of California at Davis, Dept. of Physics, Davis, CA 95616
}
\end{center}
\hspace{3in}

\begin{abstract}
We propose a mechanism for hiding the
primordial baryon asymmetry from interactions
that could wash it out. It requires the introduction of
a baryon number carrying
singlet which is in equilibrium in the early universe
and shares any existing baryon asymmetry.
It decouples from the Standard Model particles 
before all the interactions required to wash out the asymmetry
are in equilibrium ($T \simeq 10$ TeV), and decays after the
electroweak phase transition, but before nucleosynthesis.
This mechanism can conserve a baryon asymmetry in models
(a) with  $B-L = 0$, such as many $SU(5)$ GUTs,
or (b) with $B-L$ violating
interactions in thermal equilibrium, such as SUSY with
broken $R$-parity. As a result, cosmological
constraints on $R$-parity violating operators are
relaxed considerably.
\end{abstract}


Making the observed  baryon asymmetry of the universe
(BAU) \cite{BAU,K+T,EPTBAU} 
in supersymmetric theories \cite{SUSY}
with $R$-parity violation \cite{recent R-parity,RPexpt} 
can be a challenging task.
The  difficulty is that such
theories contain $B-L$ violating interactions which are
naturally in thermal equilibrium above the electroweak phase
transition, in conjunction with the
anomalous  \cite{B+L} $B+L$ violating
Standard Model processes. Together, these interactions can wash
out any BAU present in the early universe \cite{cosm1,cosm-rp}. 
The Hubble expansion rate at $T \simeq $ 100
GeV is  then so much smaller than particle interaction rates, that it is
difficult (but not impossible \cite{lowT}) to find enough perturbative 
out-of-equilibrium dynamics to (re)generate
an asymmetry. Since Supersymmetric theories
with $R$-parity violation have recently attracted
attention \cite{recent R-parity,RPexpt}, it is of
interest to consider how this problem can be avoided.
One way around it  is
to create the asymmetry at \cite{EPTBAU}
or after  the electroweak phase transition \cite{EPTBAU,lowT}.
In this letter, we follow
a  second approach, which is to assume the asymmetry
is generated
earlier, but protected by an approximate symmetry.   
We hide the primordial 
BAU in a pair of $SU(3) \times SU(2) \times U(1)$
singlets, $S$ and $\bar S$,
with unit baryon number ($B=1$) during the critical
period (ie. the time when all interactions
required to washed out any BAU are in thermal equilibrium).
As $S$ and $\bar S$ decay the BAU will be
transferred back to the Standard Model quarks.

Three ingredients are required to generate
a baryon asymmetry \cite{Sak}:
baryon number violation, $C$ and $CP$ violation, and
some out-of-equilibrium process.
These are all present in the Standard Model (SM)
at the electroweak phase transition;
 however, there is insufficient
perturbative $CP$ violation \cite{CP} in the SM, and the 
 non-perturbative $B+L$ violation  would be  in
thermal equilibrium after the transition for most 
 allowed values of the Higgs mass  \cite{Herbi} (in which case
any asymmetry produced would be destroyed). This 
suggests  that the observed baryon asymmetry
cannot be made in the Standard Model, and is
evidence for some kind of new physics.

There are numerous extensions of
the Standard Model that include viable
baryogenesis mechanisms \cite{BAU,K+T,EPTBAU}. 
It is particularly interesting that it may be
possible to
create the BAU at the
electroweak phase transition for certain regions
of parameter space in the Minimal Supersymmetric Standard
model (MSSM) \cite{MSSMBAU}.
However, we may not live in
these regions of parameter space, so  an alternative mechanism
is certainly desirable. In this letter, we
assume that the asymmetry was created before the
phase transition. Two generic and popular
mechanisms for this are
the out-of-equilibrium decay of heavy
GUT (Grand Unified Theory) particles
produced in the decay of
the inflaton \cite{GUTBAU},  or  the Affleck-Dine
mechanism  in Supersymmetry
and Supergravity \cite{A+D}.
In any case, the asymmetry produced 
is in  thermal equilibrium  in the early Universe
at temperatures above the electroweak phase transition. 
If interactions capable of washing out the asymmetry
are simultaneously present, the asymmetry will be lost. 
This can happen for an asymmetry with $B-L = 0$ within
the Standard Model, and for any asymmetry in
Supersymmetric models with sufficient
$R_p$ violation~\cite{cosm-rp}.

Recently, there has been considerable interest in
supersymmetric models with broken $R$-parity\cite{recent R-parity},
so we will
briefly review  their baryogenesis difficulties.  
$R_p$ is a multiplicative symmetry \cite{r-parity}
which assigns to each
scalar or fermionic field  in
the model the charge $(-1)^{3B + L + 2S}$,
where 
$S$ is the particle spin. 
$R_p$ conservation in supersymmetric
theories was introduced to eliminate
renormalizable $B$ and $L$ number violating
interactions, which, if present simultaneously, would induce proton
decay. Requiring $R$-parity conservation to ensure 
proton stability
may be too strong a constraint, since it is sufficient to build
models that 
eliminate either $L$ or $B$ violating interactions.
However, such models may still have problems preserving
the cosmological baryon asymmetry; if interactions
that take $B$ or $L$ to zero are in thermal
equilibrium in the presence of the anomalous $B+L$ violation,
then  any previously existing
asymmetry would be washed out \cite{cosm1, cosm-rp}.
 A primordial asymmetry with $B - L \neq 0$ 
can be protected  if there is no
perturbative $B$ violation, and  at least one lepton flavor is effectively
conserved. This constrains
the $L$ violating couplings in one generation
to be small enough that they are not in chemical  equilibrium
at the relevant temperatures.
This scenario is incompatible with our  theoretical prejudice
of relating lepton flavor violation to quark flavor violation.
Furthermore, it may be in contradiction with 
the experimental evidence for neutrino oscillations.
The main purpose of this letter is to demonstrate that
these cosmological constraints on $R_p$ violating couplings
can be circumvented by introducing new particles
that can temporarily store the baryon asymmetry.

\begin{table}[t]
$$
\begin{array}{ccccc}
\hline
\hline
{fields} 
  & SU(3)_c & SU(2)_L & U(1)_Y & B \\
\hline
{     S}  &   1 & 1 & \phantom{-}0 & \phantom{-} 1 \\
{\bar S}  &   1 & 1 & \phantom{-}0 &            -1 \\
{     T  }  &      3 & 1 & -2/3 & \phantom{-} 2/3 \\
{\bar T  }  & \bar 3 & 1 & \phantom{-} 2/3 & -2/3 \\
\hline
\end{array}
$$
\caption{
The particle content and their quantum numbers.
}
\label{p-spectrum}
\end{table}

Let us consider an extension of
the MSSM with the additional
fields and their 
quantum numbers given in table~\ref{p-spectrum}.
The most general superpotential involving these new particles
can be written as\footnote{%
The property of gauge coupling unification can be maintained by
including a additional pair of Higgs doublets with mass $m_T$.
For simplicity we neglect all interactions involving
these Higgs doublets.}
\bea
W_{\rm new} =    \lambda^{\prime} \bar T U^c D^c 
         + \lambda      T D^c S + \bar\lambda T Q Q 
+ m_T \bar T T + m_S \bar S S \,,
\label{new}
\eea
where $U^c$ and $D^c$ ($Q$) are the right (left) handed
quark fields and we assume the hierarchy $m_T \gg m_S \gsim \mz$.
For simplicity we have suppressed the flavor indices
in eq.~\ref{new}.
After integrating out the heavy fields $T$ and $\bar T$
any communication between the MSSM particles and the fields $S$ and $\bar S$
proceeds via the non-renormalizable
interaction term
\bea
W_{NR} = {[\lambda, \lambda^{\prime}] \over m_T} U^c D^c D^c S +\hc\,,
 \label{susynr}
\eea
where we have abbreviated
$[\lambda, \lambda^{\prime}] U^c D^c D^c
\equiv (\lambda_k \lambda_{i j}^{\prime} - \lambda_j \lambda_{i k}^{\prime})
 U^c_i D^c_j D^c_k$ (here, $i, j, k = 1,2,3$ are the generation indices).

This model must satisfy three requirements to protect a 
primordial baryon asymmetry. The singlet $S$ must be initially
in chemical equilibrium with the Standard Model fields,
so that $S$ and $\bar{S}$ share
the primordial BAU irrespective of its origin 
(This constraint could be relaxed by assuming
that sufficient asymmetry was generated in $S$ and $\bar S$
but we prefer to be as general as possible.)
The singlets must then decouple from
the quarks before enough interactions have come into
equilibrium to wash out all the asymmetries present.
Finally  the singlet 
must decay after the electroweak phase transition,
and sufficiently before nucleosynthesis to restore
 a homogeneous and isotropic  radiation dominated Universe
with the baryon asymmetry stored in the quarks, and 
$\Omega_o \simeq 1$.
 We will see that there is room for our model
to sit comfortably between these bounds.

The first requirement is easy to meet. The couplings
$\lambda$ and $\lambda^{\prime}$ from equation (\ref{new}) must be
sufficiently large such that  $S$ is in chemical equilibrium
with the quarks at some time
in the early Universe.  We assume that the Universe
has a reheat temperature, $T_{\rm rh} \simeq 10^{8}$ GeV,
to avoid the gravitino problem \cite{gravitino}. 
(If it is hotter, then the
lower bound we compute would be decreased.)  If the
triplet mass is less than $10^8$ GeV,  it
will be present in the thermal bath, and  $S$ will
be in chemical equilibrium with the quarks  if
 the  $T S D^c$ and $\bar T U^c D^c$ interactions are  in equilibrium
at $T \simeq m_T$, or \cite{cosm-rp}
\be
 10^{-2}  \lambda^2 m_T \,,
 10^{-2}  \lambda^{\prime 2} m_T \lappeq  H \simeq \frac{20 m_T^2}{m_{pl}}  \,,
\ee
neglecting generations and color factors.
This implies 
\be
\lambda^2, \lambda^{\prime 2} > 2 \times 10^3\frac{ m_T}{m_{pl}}  
~~~{\rm for ~} m_T \lappeq T_{\rm rh} \label{bd1a}
\ee
 or $\lambda^2, \lambda^{\prime 2} > 2 \times 10^{-13} m_T/$ TeV.

If $m_T \gg T_{\rm rh} \simeq 10^5$ TeV, then $T$ and $\bar T$
will not be thermally produced, and $S$ will have to be brought into 
chemical equilibrium  by the non-renormalizable interaction 
in eq.~(\ref{susynr}). Requiring this interaction to be in
thermal equilibrium at $T \simeq T_{\rm rh}$ gives
\be
\Gamma_{NR} \sim \frac{ [\lambda, \lambda^{\prime}]^2}{ 4 \pi m_T^2} 
\frac{T_{\rm rh}^3}{4} 
\simeq 10^{-2} \frac{[\lambda, \lambda^\prime]^2 T_{\rm rh}^3 }{m_T^2} > H \simeq 
\frac{20 T_{\rm rh}^2}{m_{pl}} \label{GNR}
\ee 
 This implies
\be
[\lambda, \lambda^\prime] >   10^{-9} \frac{ m_T}
{{\rm TeV}}  ~~~~{\rm for} ~m_T \gg T_{\rm rh} \label{bd1b}
\ee

For $m_T \gappeq T_{\rm rh} = 10^8$ GeV,
the triplets $T$ and $\bar T$
will be  present in the thermal bath with a Boltzmann suppressed
number density. The decays and inverse decays of $T$ can then
keep $S$ in chemical equilibrium with the quarks. We estimate the
rate for these processes to be of order
\be
 \Gamma \simeq \frac{  \lambda^2 m_T}{16 \pi} e^{-m_T/T_{\rm rh}} \label{G}
\ee
Requiring this to be greater than the
expansion rate at $T = T_{\rm rh}$  gives
\be
\lambda^2 > 10^{-3}  \left( \frac{{\rm TeV}}{m_T} \right) e^{m_T/T_{\rm rh}}
~~~ m_T \simeq T_{\rm rh} \label{bd1c}
\ee
 which can be used to join the constraints from eqs.~(\ref{bd1a}) and
(\ref{bd1b}) at $m_T \simeq T_{\rm rh}$.

The next step is to determine the lower bound on
$[\lambda, \lambda^\prime]/m_T$ 
 from requiring that $S$ and $\bar{S}$ decouple from the MSSM
while there are still asymmetries present in the plasma. 
Thus, we must first determine 
the temperature at which the asymmetry gets washed out.
At very high temperatures in the early Universe, most
Yukawa couplings are too weak to be in thermal equilibrium;
as the temperature drops,
more and more of them come into equilibrium. For the electron
Yukawa coupling this happens at
$T \simeq 1-10$ TeV.
An asymmetry in
the right-handed electron, $e_R$,
carries net electric charge, which has to be compensated by
asymmetries carried by other particles to ensure charge
neutrality of the Universe.
Thus, in the MSSM the baryon asymmetry will remain until this
temperature \cite{CDEO3}.

In our model with broken $R_p$ there are other couplings
involving $e_R$. Hence, the temperature at which the baryon asymmetry
gets washed out depends on the strength of the lepton flavor
violating interactions.
Let
\be
W_{\rs} = 
\mu_i \bar{H}  L_i + \half y_{i j k}^L L_i L_j E^c_k + 
y_{i a b}^D L_i Q_a D^c_b + \half \bar y^D_{i j k} U^c_i D^c_j D^c_k
\label{w r-parity}
\ee
be the $R$-parity violating part of the superpotential.
Clearly, it is quite complicated to determine the region
in parameter space where $e_R$ is out of chemical
equilibrium at $T\simeq \TB \equiv 10$~TeV
due to the large number of parameters
and because any constraint on an individual
coupling has to be formulated in a particular
basis.
A sufficient but not necessary condition is to assume
that the $R_p$ violating
couplings involving  the $i$th lepton generation are
smaller than the Yukawa coupling of
that generation.
In the MSSM thermal mass eigenstate basis
this condition on the parameters reads
\bea
\bar y^D_{i j k} &=& \hbox{arbitrary}\,,\nonumber\\
{\mu_i \over \mu_0} y^D_{3 3} \, ,
y_{i a b}^D, \, y_{i j k}^L&\lsim& {g m_\ell \over \sqrt{2} m_{W}}\,,
\qquad\hbox{for}~i, j, k \leq \ell\,, 
\label{r-parity constraints}
\eea
where $m_1<m_2<m_3$ denote the three charged lepton masses
and $y^D_{3 3}$ is the bottom Yukawa \cite{DE1}.
In this case, the  baryon
asymmetry will generically be preserved
until the electron Yukawa comes into chemical equilibrium
at $T \simeq 1-10$ TeV. We therefore require $S$ to be
out of chemical equilibrium by $\TB \simeq 10$ TeV. 

We get a lower bound on $[\lambda, \lambda^\prime]/m_T$ by
requiring that the dimension 5 operator of eq.~(\ref{susynr})
to be out of equilibrium 
before the temperature drops below $\TB$. Requiring the rate
$\Gamma_{NR}$ from eq.~(\ref{GNR}) 
to be less than the Hubble expansion
at $\TB$ gives
\be
\Lambda_{NR} \equiv 
{[\lambda, \lambda^{\prime}]\over m_T} 
< 9 \times  10^{-8} \tev^{-1} \,.
\label{bd2}
\ee
This constraint applies to all the coupling constant combinations
$[\lambda, \lambda^{\prime}]_{i j k}$, because the singlet
$S$ has to be out of chemical equilibrium with all
the MSSM particles.

The constraint of eq.~(\ref{bd2}) was estimated from the zero
temperature scattering cross-section for processes 
 where a heavy off-shell $T$  is exchanged. At finite temperature
in the early Universe, there will also be some number of
$T$ particles present in the thermal bath, and their interaction
rates below $\TB$ must also be small enough to
keep $S$ out of chemical equilibrium with the quarks.
Using the estimate [eq.~(\ref{G})] for  the
interaction rate of $S$ with the quarks
via decays and inverse decays of $T$s present in the
thermal bath, and requiring this to be less than the expansion rate
at $\TB$ gives
\be
\lambda^2 <  8 \times 10^{-12} \left( \frac{{\rm TeV}}{m_T}
\right) e^{m_T/\TB} \label{bd3}
\ee

\begin{figure}
\vspace*{10pt}
\vspace*{4.truein}             
\includegraphics{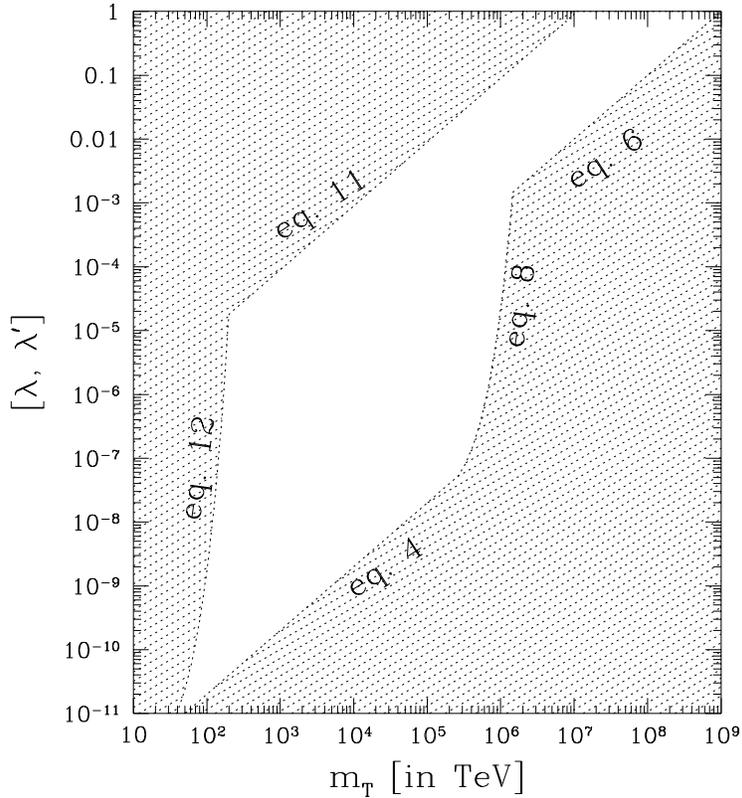}
\caption{ The area in the \protect\( 
[\lambda, \lambda'] \protect\) --
\protect\( m_T\protect\) plane allowed by the requirements that  the singlet
$S$ be in equilibrium early in the Universe (this rules out
area below the lower diagonal curve), and out of equilibrium
by $T \simeq 10$ TeV (this rules out the area above
the upper line, and to the left of the left-hand curve).
We set \protect\( \lambda^2 = \lambda^{\prime 2} =
[\lambda, \lambda'] \protect\).}
\label{fig1}
\end{figure}

In figure~\ref{fig1}, we plot the bounds on $[\lambda, \lambda^{\prime}]$
as a function of $m_T$.
For simplicity we assume
$[\lambda, \lambda^{\prime}] = \lambda^2 = \lambda^{\prime 2}$.
Note, that the antisymmetrization of $[\lambda, \lambda^{\prime}]$
below eq.~(\ref{susynr}) only projects out the flavor changing components.
In a realistic model with approximate flavor symmetries
these are expected to be suppressed.
Hence, the lower bounds on $[\lambda, \lambda^{\prime}]$
[which are actually constraints on
$\lambda^2$ and $\lambda^{\prime 2}$ from
eq.~(\ref{bd1a}) and eq.~(\ref{bd1c})] are very conservative.
They are obtained by requiring that the singlet
be initially in equilibrium with the MSSM, and
then out of equilibrium by $T \simeq \TB = 10 $ TeV
[eq.~(\ref{bd2})]. 
In a realistic three generation model, 
the  lower line  corresponding to eqs.~(\ref{bd1a}), (\ref{bd1b}) and (\ref{bd1c}), 
applies to the largest component of $[\lambda, \lambda^\prime]_{i j k},
|\lambda_i|^2$ or $|\lambda'_{jk}|^2$ that
couples $S$ to quarks $U^c_j$, $D^c_i$ or $D^c_k$,  one of which carries 
a primordial asymmetry.
We expect that in a generic
GUT baryogenesis model, some asymmetry would be created 
in the second or third generation leptons or quarks. At
$T \simeq 10^8$ GeV, the sphalerons and these particles
Yukawas are in equilibrium,  so the asymmetry would be
shared among them. If $S$ couples sufficiently strongly to
at least one second or third generation particle, it will
also acquire an asymmetry.\footnote{%
This constraint does not need to be respected
if a GUT-scale asymmetry of
the right magnitude was created directly in $S$.}

Masses to the left of  the left line are ruled out,
because the particles $T$
would be present in the thermal bath in sufficient numbers
to mediate $S$ to  quark  transitions [eq.~(\ref{bd3})].
This bound applies to
all the $\lambda_j$, where $j$ is a quark generation index.
Clearly, there is a substantial region in the 
$m_T$---$[\lambda, \lambda^{\prime}]$ plane consistent with all
the bounds in eqs.~(\ref{bd1a}),  (\ref{bd1b}) and  (\ref{bd1c})--(\ref{bd3}).

We now turn to the third requirement which guarantees that
the fermionic and scalar singlets decay in such a way as
to preserve
a BAU without disrupting primordial nucleosynthesis or
other cosmological observations. 
Thus, we have to look at the mass spectrum and decay pattern of
$S$ and $\bar S$.
We assume that SUSY is broken explicitly by
soft SUSY breaking terms \cite{gira}.
Furthermore, we keep all relevant soft SUSY breaking terms 
degenerate with the exception of the mass of the lightest SUSY partner
(LSP), $m_{LSP} \ll \msusy$.
The mass spectrum then looks as follows: 
there are two complex scalars
$S_{1,2} = (S \pm \bar S^*)/\sqrt{2}$
with mass
$m_{S_{1,2}}^2 = m_S^2+\msusy^2\pm B m_S$
($B$ is the soft SUSY breaking term multiplying the bilinear
term in $W$; in our numerical work we set $B = \msusy$) 
and one four component Dirac fermion
$s = (\psi_S, \bar \psi_{\bar S})$
with mass $m_S$.

Now we have to find the range of parameters where
the longest-lived singlet, $s$ or $S_{i}$ ($1= 1,2$),
decays after the EPT 
(so that the baryon asymmetry it carries is not 
washed out by the sphaleron transitions),  but
sufficiently before nucleosynthesis to not disrupt
the process of light element formation. The constraints
on low temperature baryogenesis models from nucleosynthesis
were studied in \cite{nucl}, where it was shown that
nucleosynthesis will proceed in the standard fashion
if the post-baryogenesis Universe reheats to
look like a standard Big Bang model with $T \gappeq 3$ MeV.
We take ``sufficiently
before nucleosynthesis'' to mean 
that the (instantaneous) reheat temperature
after the decays of $S_i$ and $s$ should exceed 100 MeV; this means
that all but $ e^{-1000}$ of the singlets will have decayed
by $T \simeq 3$ MeV.  If the Universe is
radiation dominated when $S_i$ and $s$ decay, then we
require
\be
10^{-10} {\rm sec} \simeq H^{-1}(T_{EPT}) <  \Gamma_S^{-1}  < H^{-1} (T \simeq 
 100 ~ {\rm MeV})  \simeq 10^{-4} ~ {\rm sec} \label{bd4a}
\ee
where $\Gamma_S \equiv {\rm min} ( \Gamma_{S_i}, \Gamma_s)$ is
the decay rate of the particles $S_i$ and $s$ with the
longest life-time.  For a large range of 
parameter space, $S_i$ or $s$ will dominate the
energy density of the Universe before they decay. In this
case, we want  $\rho_S$, the energy density in $S_i$ (or
$s$ if they decay last) to be greater than $\rho_{rad} (T \simeq 100
~ {\rm MeV}) = g_{eff}(T) \pi^2 T^4/30$.  This means $\Gamma_S > H(T \simeq 
100~ {\rm MeV})$, as above.

>From the widths for the dominant decay modes
\bea
\Gamma(S_i\rightarrow 2q+\tilde q)
 &\simeq& \kappa m_{S_i}^3 \Lambda_{NR}^2 f(x)\,,
\nonumber\\
\Gamma(S_i\rightarrow 3q+LSP)
 &\simeq& {\alpha_{\rm em}\over 4 \pi}
\kappa m_{S_i}^3 \Lambda_{NR}^2\,,
\nonumber\\
\Gamma(s\rightarrow 2\tilde q+q)
 &\simeq& \kappa m_{S}^3 \Lambda_{NR}^2 g(x)\,,
\nonumber\\
\Gamma(s\rightarrow 3q)
 &\simeq& \left({\alpha_{\rm s}\over  3\pi}\right)^2
\kappa m_{S}^3  \Lambda_{NR}^2
{\rm min} \{1, x^{-1}\}\,,
\label{Gd}
\eea
we obtain
\be
\Gamma_S = \kappa \Lambda_{NR}^{2}{\rm min}\left\{
m_{S_i}^{3} \left[f(\tilde x_i) + {\alpha_{\rm em}\over 4 \pi}\right]\,,
m_S^{3} \left[g(x) + \left({\alpha_{\rm s}\over 4 \pi}\right)^2
\, {\rm min}\{1, x^{-1}\}\right]
\right\}\,.
\ee
Here, we set the fine structure constant $\alpha_{\rm em} = 1/137$
the strong coupling $\alpha_s = 0.11$ and $\kappa = 1/(6144 \pi^3)$.
Furthermore, we have defined
$\tilde x_i = (m_{\tilde q}/m_{S_i})^2$,
$x = (m_{\tilde q}/m_{S})^2$  and
\bea
f(x) &=& 6 x (1+x) \ln(x)+(1-x) (1+10 x+x^2) \,,\nonumber\\
g(x) &=& 3(2 x^2-x-2 x^3)\left\{\half \ln(4 x)-\ln(1-\sqrt{1-4 x})
     -\half\ln\left[ {(1-3 x)-(1-x)\sqrt{1-4 x}
                 \over(1-3 x)+(1-x)\sqrt{1-4 x}}\right]\right\}\nonumber\\
&&+(5 x-6 x^2+1)\sqrt{1-4 x}\,,
\eea
The decay $s \rightarrow 3q$ proceeds at the one-loop level
where we have again assumed that squarks and gluinos are
mass-degenerate.

\begin{figure}
\vspace*{10pt}
\vspace*{4.truein}             
\includegraphics{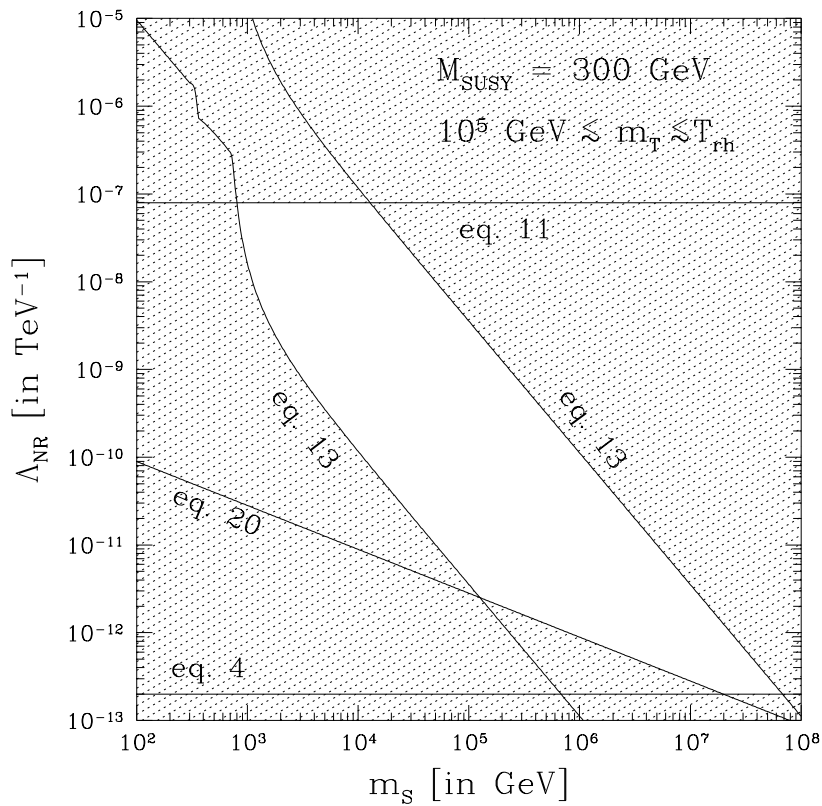}
\caption{ 
 The area in  $\Lambda_{NR}$  (in  TeV$^{-1}$) vs
$m_S$ space allowed after requiring 1) that $S$ be in equilibrium
early (rules out below the lower horizontal line) and out
of equilibrium by 10 TeV (rules out above the
upper  horizontal line), 2) that the fermion $s$ decay
after the electroweak phase transition (excludes to the right of the right
jagged curve), but before nucleosynthesis (excludes
to the left of upper left curve), and 3) that
the decay products of $s$ and $S_i$ have time
to thermalize before nucleosynthesis (this rules
out below the lower left curve). These bounds
assume $\msusy = 300 $ GeV.}
\label{fig2}
\end{figure}

In figure~\ref{fig2}, we plot  the constraints 
in the $m_S$---$\Lambda_{NR}$ plane for
$\msusy = 300$ GeV.  The two horizontal
lines correspond to bounds of eqs.~(\ref{bd1b}) and (\ref{bd2});
the area between the two lines is allowed. 
The constraints in eq.~(\ref{bd4a}) rule out anything outside the
diagonal curves.
In the upper right-hand corner $S_{i}$ and $s$
would decay before the EPT. 
The lower left-hand corner is ruled out
by our conservative requirement that  only
$ \simeq 10^{-400}$ of the $s$ and $S_{i}$ should be left to decay
during nucleosynthesis.

Finally, we must also require that the
decay products of $s$ and $S_{i}$ thermalize efficiently
before nucleosynthesis, so that they
do not disassociate any light elements. With a baryon
to photon ratio of $\eta \simeq  2 - 4 \times 10^{-10}$ \cite{BBN}
having one overly energetic particle in $10^{10}$
means that there is one dangerous particle per baryon. 
Thus, we not only have to guarantee that the singlets have decayed
sufficiently before nucleosynthesis [eq.~(\ref{bd4a})]
but also that their decay 
products have sufficiently thermalized by $T \simeq 3 $ MeV.
Thus, let us consider the universe shortly after
the decay of $s$ where we expect a soup of squarks,
quarks or nucleons, 
and LSPs with energies of order $m_S$.  The
squarks will rapidly decay.  The quarks will
scatter and emit lower energy quarks and gluons at
a rate of order
\be
\Gamma_{scat} \simeq \frac{\alpha_{\rm s}^3}{m_S^2} n_s \label{scatt}
\ee
where $n_s$ was the pre-decay number density of $s$:
\be
n_s \sim \frac{\rho_{dec}}{m_S}  = \frac{3 m_{pl}^2 \Gamma_{S}^2}
{8 \pi}
\ee   
If the particles to be thermalized are nucleons, the scattering
cross-section is much larger, and we expect the time-scale we compute
for quark thermalizations to be more than long enough.
 
The time-scale to scatter once is of order the thermalization
time-scale \cite{K+T}, so  we require  that 
$\Gamma_{scat}$ (scaled to $ T \simeq 3 $ MeV)  be  more than
a factor of a hundred larger than  the expansion rate at $T_n \simeq 3$ MeV:
\be
\frac{\alpha_{\rm s}^3}{m_S^2} n_s \frac{R_d}{R_n} > 10^2 \times H (T_n)
 \label{scat}
\ee
where $R_d ~ (R_n$) is the scale factor when
the singlet decays (at nucleosynthesis).
This means that at most  a fraction 
$\lappeq e ^{-100}$ of the decay products could
be unthermalized at $ T \simeq 3$ MeV. 
Expressing $H$ and the ratio of scale factors in
terms of the energy density at decay [or equivalently,
the decay rate of eq.~(\ref{Gd})], we get the bound
\be
\Lambda_{NR} = \frac{ [\lambda,  \bar{\lambda}]}{m_T} >  3 \times 10^{-11}  
\sqrt{ \frac{ {\rm TeV}}{m_S}} ~
{\rm TeV} ^{-1}
\ee
This is the lower diagonal line on the left in figures~\ref{fig2}.
As expected, the thermalization bound is only relevant
for weakly interacting (small $\Lambda_{NR}$) heavy singlets.

The decay of $s$ can produce particles
other than quarks, who must also decay and/or
thermalize. If these are Standard Model
fermions or gauge bosons, this should be no
problem.  If they are heavy superpartners, 
they will also decay soon enough (the Universe at
$T \simeq 100 $ MeV is $\simeq 3 \times 10^{-5}$ seconds
old). The only potential problem is the LSP, which
in our model is assumed to be unstable and has a life-time of
\be
\tau_{LSP} \simeq 4 \times \frac{10^{-18} s}{\lambda_{\rs}^2}
\left( \frac{ \msusy}{300~\gev} \right)^4 
\left( \frac{50~\gev}{m_{LSP}}\right)^5\,,
\ee
where $\lambda_{\rs}$ stands generically
for the dominant $R$-parity violating
Yukawa coupling in eq.~(\ref{w r-parity}).
Requiring that $\tau_{LSP}\lsim H^{-1}(100 ~\Mev)$ yields
the constraint
\be
\lambda_{\rs} \gsim 2 \times 10^{-7}
\left( \frac{ \msusy}{300~\gev} \right)^2
\left( \frac{50~\gev}{m_{LSP}}\right)^{5/2} \,,
\label{pr-tau(lsp)}
\ee
which may not be necessary but is certainly sufficient.


We note that for most of the allowed parameter
space for this model, the singlet $s$ will dominate
the energy density of the Universe before it decays, and 
can therefore generate substantial entropy.
 This means that the baryon excess
stored in the singlet must be somewhat larger than
in the standard scenario. This is straightforward to quantify: if
\be
\epsilon \simeq \frac{n_s - n_{\bar{s}}}{n_s + n_{\bar{s}} }\simeq 
\frac{\mu_s}{T} \,,
\ee
where $\mu_s$ is the singlet chemical potential
when it was in chemical equilibrium, and
$n_s$ is the sum of the
number densities of $S_{1}$, $S_{2}$ and $s$, then the baryon
to entropy ratio today is
\be
\frac{n_B}{\sigma} \simeq \epsilon \frac{n_s}{\sigma}
\ee
where $\sigma$ is the entropy density (to avoid confusion
with the singlet fermion $s$).
If $s$ dominates the energy density of the Universe
before it decays, then \cite{K+T} 
\be
\frac{n_s}{\sigma} \simeq \frac{ m_s}{T_{\rm rh, S}}
\ee
where $T_{\rm rh, S}$ is the temperature at which the Universe
becomes radiation-dominated after the singlets decay (``reheat 
temperature''). We assumed
this was at least 100 MeV, so for singlet masses
in the TeV range, this gives $n_B/\sigma \simeq 10^{-4} \epsilon$,
or $\epsilon \simeq 10^{-6}$. This is not excessively large.

A mechanism similar to the one we have outlined
here can be used to protect an asymmetry with
$B-L = 0$ in the Standard Model. The difference
in this case,  is that
the non-renormalizable operator
induced by the triplet is suppressed  with respect to
eq.~(\ref{susynr})
by an extra power of $m_T$: the triplet is a  scalar, and
 generates the following four fermion operator:
\be
V = \frac{ [\lambda, \lambda^\prime] }{m_T^2} s u d d +\hc
\ee
For $m_T$ less than  the reheat temperature
after inflation, and for  relatively large
values of $\lambda$ and $\bar{\lambda}$ ($\gappeq 10^{-2}$),
the non-SUSY version can satisfy the same
constraints as its supersymmetric counter-part.

To summarize, we have presented a mechanism to protect
the primordial baryon asymmetry from being washed out before the electroweak
phase transition.
A mechanism of this kind is necessary in the Standard Model if $B-L = 0$
or in supersymmetric models with $R$-parity violation.
Our mechanism is quite economical in that it only requires
the existence of a pair of singlets and a pair of Higgs triplets and
no exotic representations. 
It is also very generic in that it works in a sizable region of the
parameter space and does not require any additional assumptions about the
generation of the primordial baryon asymmetry.
Some mild constraints on $R$-parity violating parameters remain:
the LSP has to decay sufficiently before nucleosynthesis
[eq.~(\ref{pr-tau(lsp)})]; the asymmetry in $e_R$ 
should not be washed out before the singlet $S$ decouples from the
thermal bath
(a sufficient but not necessary set of constraints
is presented in eq.~(\ref{r-parity constraints})].
Note, that the $B$ violating couplings are unconstrained by the 
baryon asymmetry
in our model.

\subsection*{acknowledgements}
We would like to thank G. Raffelt and H. Dreiner for useful conversations.
One of us (RH) was supported in parts by the DOE under
Grants No. DE-FG03-91-ER40674 and by the
Davis Institute for High Energy Physics.


\begin{thebibliography}{222222}
\bibitem{BAU} 
  A.D. Dolgov, {\it Phys.Rept.} {\bf 222} (1992) 309;
\bibitem{K+T} 
   E. Kolb, M. Turner, {\it The Early Universe}, Addison-Wesley (1990).
\bibitem{EPTBAU}  see for instance, 
  M. Dine {\it  Nucl.Phys.Proc.Suppl.} {\bf 37A} (1994) 127-136 
  (hep-ph/9402252);
  A.G. Cohen, D.B. Kaplan, A.E. Nelson 
  {\it Ann. Rev. Nucl. Part. Sci.} {\bf 43} (1993) 27,  hep-ph/9302210. 

\bibitem{SUSY}
for a review, see, \eg, H.P. Nilles, \PREP {110}{1}{1984};
H.E. Haber and G.L. Kane, \PREP {117}{75}{1985};
R. Barbieri, Riv. Nuovo Cimento {\bf 11}\rm , 1 (1988).


\bibitem{recent R-parity}
V. Barger, M.S. Berger, P. Ohmann, R.J.N. Phillips, \PRD{50}{4299}{1994};
T. Banks, Y. Grossman, E. Nardi and Y. Nir, \PRD{52}{5319}{1995};
V. Barger, W.-Y. Keung and R.J.N. Phillips, \PLB{364}{27}{1995};
M. Nowakowski and A. Pilaftsis, \NPB{461}{19}{1996};
R. Hempfling, MPI-PHT-95-59, hep-ph/9511288 (Nov. 1995)
{\sl Nucl. Phys.} {\bf B} to appear;
H. Dreiner and Heath Pois, ETH-TH-95-30, hep-ph/9511444 (Nov. 1995);
V. Barger, M.S. Berger, R.J.N. Phillips and T. Wohrmann,
\PRD{53}{6407}{1996};
N.V. Krasnikov, {\it JETP Lett.} {\bf 63} 503, 1996;
B. de Carlos and P.L. White, SUSX-TH-96-003, hep-ph/9602381
(Feb. 1996), {\it Phys. Rev.} {\bf D} to appear;
K. Tamvakis, CERN-TH-96-96, hep-ph/9604343 (Apr 1996);
A. Bartl, W. Porod, M.A. Garcia-Jareno, M.B. Magro, J.W.F.
Valle, W. Majerotto, FTUV-96-29, hep-ph/9606256, (Jun 1996);
F.M. Borzumati, Y. Grossman, E. Nardi and Y. Nir, 
WIS-96-21-PH, hep-ph/9606251 (May 1996);
By M. Bastero-Gil, B. Brahmachari and R.N.
Mohapatra, hep-ph/9606447, (Jun 1996);
J.C. Romao, A. Ioannisian, J.W.F. Valle, hep-ph/9607401 (July 1996).

\bibitem{RPexpt}
G. Bhattacharyya, D. Choudhury and K. Sridhar, \PLB{355}{193}{1995};
C.E. Carlson, Probir Roy and Marc Sher, \PLB{357}{99}{1995};
G. Bhattacharyya, Amitava Raychaudhuri, \PLB{374}{93}{1996};
M. Hirsch, H.V. Klapdor-Kleingrothaus and S.G. Kovalenko, \PRL{75}{17}{1995};
A. Yu. Smirnov and F. Vissani, IC-96-16, hep-ph/9601387
(January 1996);
R. Adhikari and B. Mukhopadhyaya, \PLB{378}{342}{1996};
D. Choudhury and P. Roy, \PLB{378}{153}{1996};
M. Chaichian and K. Huitu, HU-SEFT-R-1996-09
and hep-ph/9603412 (March 1996);
H. Nunokawa, A. Rossi and J.W.F. Valle,
FTUV-96-33, hep-ph/9606445, (June 1996);



\bibitem{B+L} 
  G. 't Hooft, {\it Phys.Rev.Lett.} {\bf 37} (1976) 8;
  F.R. Klinkhamer, N.S. Manton, {\it Phys.Rev.} {\bf D30} (1984) 2212;
  P. Arnold, L. McLerran, {\it  Phys.Rev.} {\bf D36} (1987) 581.
\bibitem{cosm1} M.Fukugita, T.Yanagida, {\it Phys. Rev.} {\bf D 42} (1990) 
  1285; 
  A.E. Nelson, S.M. Barr, {\it Phys. Lett.} {\bf B 246} (1990) 141;
  W. Fischler, G.F. Giudice, R.G. Leigh, S.Paban, 
  {\it Phys. Lett.} {\bf B 258} (1991) 45;
  W. Buchm\"uller, T. Yanagida,  {\it Phys. Lett.} {\bf B302} (1993) 240.
\bibitem{cosm-rp}
  B.A. Campbell, S.Davidson, J. Ellis, K.A. Olive, {\it Phys. Lett.} {\bf
  B 256} (1991) 457, {\it Astroparticle Phys.} {\bf 1} (1992) 77; 
  H. Dreiner, G.G. Ross, {\it  Nucl. Phys.} {\bf B410} (1993) 188. 
\bibitem{lowT} 
  S.Dimopoulos, L.J. Hall, {\it Phys. Lett.} {\bf B 196} (1987) 135; 
  J. Cline, S. Raby, {\it Phys. Rev.} {\bf D 43} 
  (1991) 1781; 
  R. J. Scherrer, J. Cline, S. Raby, D. Seckel,
  {\it Phys.Rev.} {\bf D44} (1991) 3760;
  A. Masiero, A. Riotto, {\it Phys.Lett.} {\bf B289} (1992) 73. 
\bibitem{Sak}  
  A.D. Sakharov,   {\it JETP Lett.} {\bf 5} (1967) 24. 
\bibitem{CP} 
  M.B. Gavela, M. Lozano, J. Orloff, O. Pene, 
  {\it Nucl.Phys.} {\bf B430} (1994) 345; 
  M.B. Gavela, P. Hernandez, J. Orloff, O Pene
  C. Quimbay, {\it Nucl.Phys.} {\bf B430} (1994) 382.
\bibitem{Herbi}D. Diakonov, M. Polyakov, P. Sieber, J. Schaldach, K. Goeke,
  {\it Phys. Rev.} {\bf D 53} (1996) 3366. 
\bibitem{MSSMBAU}
  M. Carena, M. Quiros, C.E.M. Wagner, CERN-TH-96-30, hep-ph/9603420; 
  D. Delepine, J.M. Gerard, R. Gonzalez Felipe, J. Weyers, UCL-IPT-96-05, 
  hep-ph/9604440; 
  J. Cline, K. Kainulainen, hep-ph/9605235.
\bibitem{GUTBAU} see, for instance,
  E. W. Kolb, S. Wolfram, {\it  Nucl.Phys.} 
  {\bf B172} (1980) 224;  erratum-{\it ibid.} {\bf B195} (1982) 542;
  J.N. Fry, K. A. Olive, M. S. Turner, {\it Phys.Rev.} {\bf D22} 
  (1980) 2977, {\it ibid} {\bf D22} (1980) 2953. 
\bibitem{A+D} see, for instance, 
  I. Affleck, M.Dine, {Nucl. Phys.} {\bf B 249}
  (1985) 361 ; 
  J. Ellis , K. Enqvist, D.V. Nanopoulos, K.A. Olive, 
  {\it Phys. Lett.} {\bf  B191} (1987) 343;
  M.Dine, L. Randall, S. Thomas, {\it Phys. Rev. Lett.}
  {\bf 75} (1995) 398;
  {\it ibid.} {\it Nucl. Phys.} {\bf B458} (1996) 291;
  M. Gaillard, H. Murayama, K.A. Olive, {\it Phys.Lett.} {\bf B355}
  (1995) 71; 
  G. Anderson, A. Linde, A. Riotto, hep-ph/9606416.

\bibitem{r-parity}
G. Farrar and P. Fayet, \PLB{76}{575}{1978};
G. Farrar and S. Weinberg, \PRD{27}{2732}{1983}.

\bibitem{gravitino}see, for instance: \\
 S.Weinberg, {\it Phys. Rev. Lett.} {\bf 48} (1982) 1303; 
 J. Ellis, J.E. Kim, D.V. Nanopoulos {\it Phys. Lett.} {\bf B 145} 
 (1984) 181;
 M. Kawasaki, T.Moroi, {\it Prog. Theor. Phys.} {\bf 93} (1995) 879;
 H. Fujisaki, K. Kumekawa, M. Yamaguchi, M. Yoshimura, {\it Phys. Rev.}
 {\bf D 54} (1996) 2494.
  

\bibitem{CDEO3}  
  B.A. Campbell, S.Davidson, J. Ellis, K.A. Olive,
  {\it Phys. Lett.} {\bf B 297 } (1992) 118. 

\bibitem{DE1} S.Davidson, J. Ellis, 
 CERN-TH/96-258, CfPA 96-th-20, MPI-PhT-96-68.

\bibitem{gira}
L. Girardello and M.T. Grisaru, \NPB{194}{65}{1982}.

\bibitem{nucl} P. Delbourgo-Salvador, P. Salati, J. Audouze,
  {\it Phys. Lett. } {\bf B 276} (1992) 115. 


\bibitem{BBN} for a recent analysis, see, for instance,
  C. J. Copi, D. N. Schramm, M. S. Turner, astro-ph/9606059;
  B. D. Fields, K. Kainulainen, K. A. Olive, D. Thomas,  astro-ph/9603009;
  for a review, see, for instance,
  G. Steigman,   to be published in {\it  Nucl. Phys. Proc. Suppl.},
  astro-ph/9602029. 



\end{thebibliography}
\end{document}
\bye